\documentclass{mpe_report}
\usepackage{psfig}
\usepackage{graphicx}

\begin{document}
\title{A multi-coloured survey of NGC 253 with XMM-Newton}
\author{R. Barnard\inst{1} \and L. Shaw Greening\inst{1} \and U. Kolb\inst{1}}  
\institute{The Open University}
\maketitle

\begin{abstract}
There is a large body of work that has used the excellent Chandra observations of nearby galaxies with neglible low mass X-ray binary (LMXB) populations. This has culminated in a ``Universal'' X-ray luminosity function (XLF) for high mass X-ray binaries  (HMXBs). However,  a number of methods have been used to convert from source intensities to luminosities when creating these XLFs. We have taken advantage of the XMM-Newton observations of the nearby starbursting spiral galaxy NGC 253 to test some of these methods. We find the luminosities derived from these various methods to vary by a factor of $\sim$3. We also find the most influential factor in the conversion from intensity to luminosity to be the absorption. We therefore conclude that a more consistent approach is required for determining the true Universal XLF for HMXBs. Ideally, this would involve individual spectral fitting of each X-ray source. Certainly, the line-of-sight absorption should be determined from the observations rather than assuming Galactic absorption. We find the best approach for obtaining an XLF from low-count data to be the splitting of the X-ray sources into two or more intensity intervals, and obtaining a conversion from intensity to flux for each group from spectral modelling of the summed spectrum of that group.

\end{abstract}

\section{introduction}
The X-ray populations of external galaxies have been well studied for the last $\sim$20 years. Historically, studies of the individual sources have been severely limited by low count rates and signal to noise, and several methods have been used to derive the X-ray luminosity of a source from its intensity.
Grimm et al. (2003, hereafter known as G03) used Chandra and ASCA surveys of nearby starburst galaxies, along with ASCA, MIR-KVANT/TTM and RXTE/ASM observations of HMXBs in our Galaxy and the Magellanic Clouds to obtain a correlation between the X-ray properties of HMXBs and the star formation rate (SFR) of their host galaxies. They chose their sample of galaxies to have sufficiently high SFR to total mass ratios so that their X-ray populations would be dominated by HMXBs, with negligible LMXB contributions.  
G03 used published Chandra X-ray luminosity functions (XLFs), scaled  assuming the Hubble constant to be 70 km s$^{-1}$ Mpc$^{-1}$. They found the XLFs of these galaxies to be strikingly similar, when normalised by the SFR of the galaxy, and proposed a universal HMXB XLF. Additionally,  they found that the number of sources with 2--10 keV luminosities $>$2$\times$10$^{38}$ erg s$^{-1}$ to be proportional to SFR$^{1.06\pm0.07}$. Furthermore they discovered a linear relation between the total HMXB X-ray flux of a galaxy and its SFR, for SFRs $\ga$4 M$_{\odot}$ yr$^{-1}$.

Several different methods were used to convert from X-ray intensity to luminosity when creating the XLFs for galaxies in the G03 sample. Some XLFs were derived  assuming  a standard X-ray binary emission model (a power law with spectral index, $\Gamma$, $\sim$1.7 or a 5 keV bremsstrahlung) with Galactic line-of sight absorption (e.g. Zezas et al., 2002; Soria \& Kong, 2002). Others used the X-ray colours to estimate the emission spectrum, using Galactic absorption (e.g. Eracleous et al., 2002), or deriving the absorption from the colours also (e.g. Lira et al., 2002). Also, some XLFs were derived using best fit spectra to individual bright sources (Smith \& Wilson, 2001), or to the stacked X-ray population (Roberts et al., 2002).

XMM-Newton (XMM) is the most sensitive X-ray imaging telescope in the 0.3--10 keV band. We can therefore use deep XMM observations of nearby galaxies to glean the accuracy of these methods by comparing their resulting XLFs with the XLF derived from freely fitting each X-ray source. To do this, we chose XMM-Newton observations of NGC 253.
NGC 253 is a star-bursting spiral  galaxy in the Sculptor  Group at a distance of $\sim$4 Mpc; it is almost edge on, with a D$_{25}$ region of $\sim$25$'$$\times$7$'$. We fully describe our analysis of the observations in Barnard et al. (2007, MNRAS submitted,  hereafter known as B07), and concentrate here on the 2003, 110 ks XMM observation of NGC253.

\section{Analysis}
\begin{figure}
\begin{centering}
\resizebox{\hsize}{!}{\includegraphics[angle=0,scale=0.3]{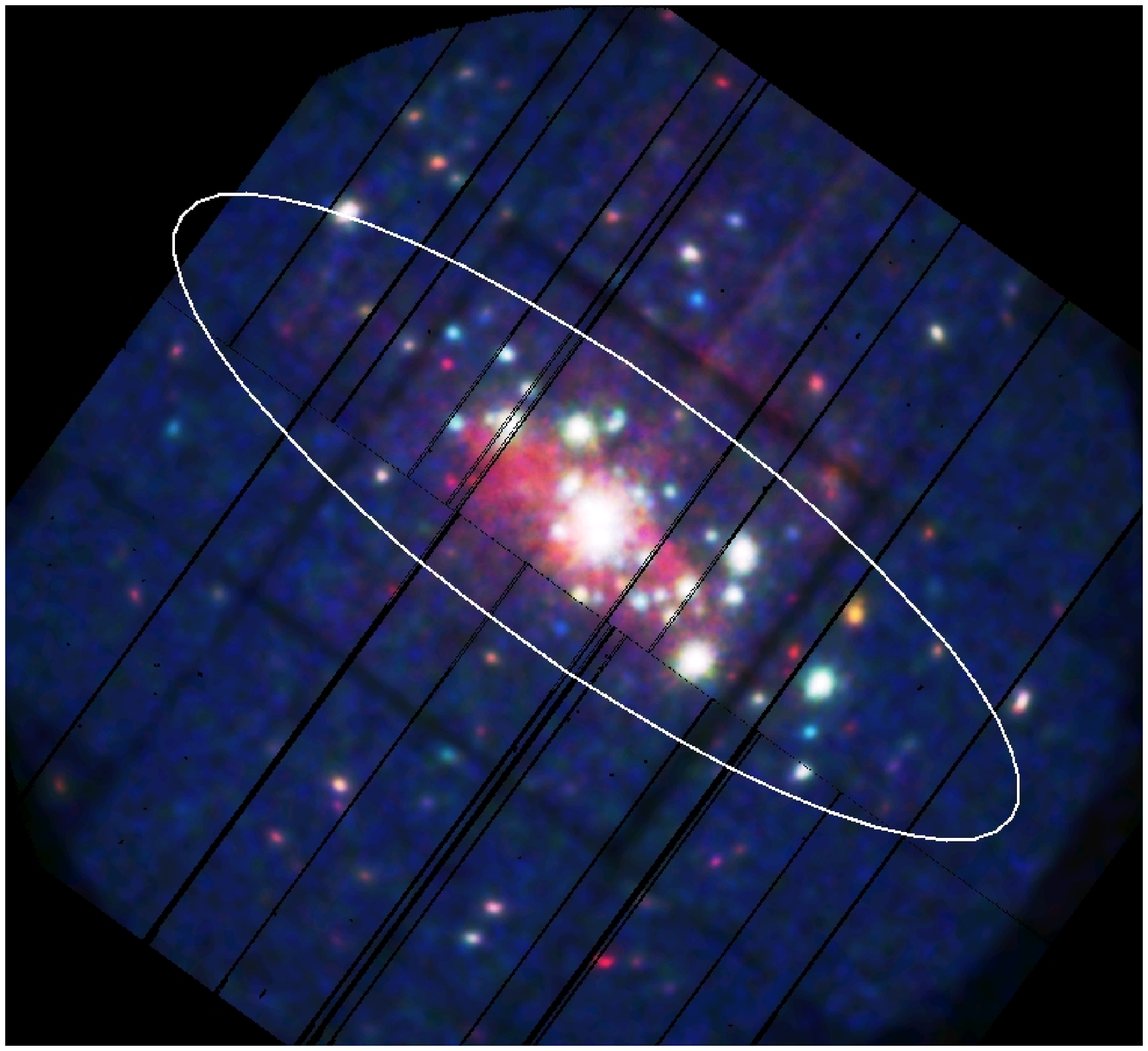}}
\resizebox{\hsize}{!}{\includegraphics[angle=0,scale=0.5]{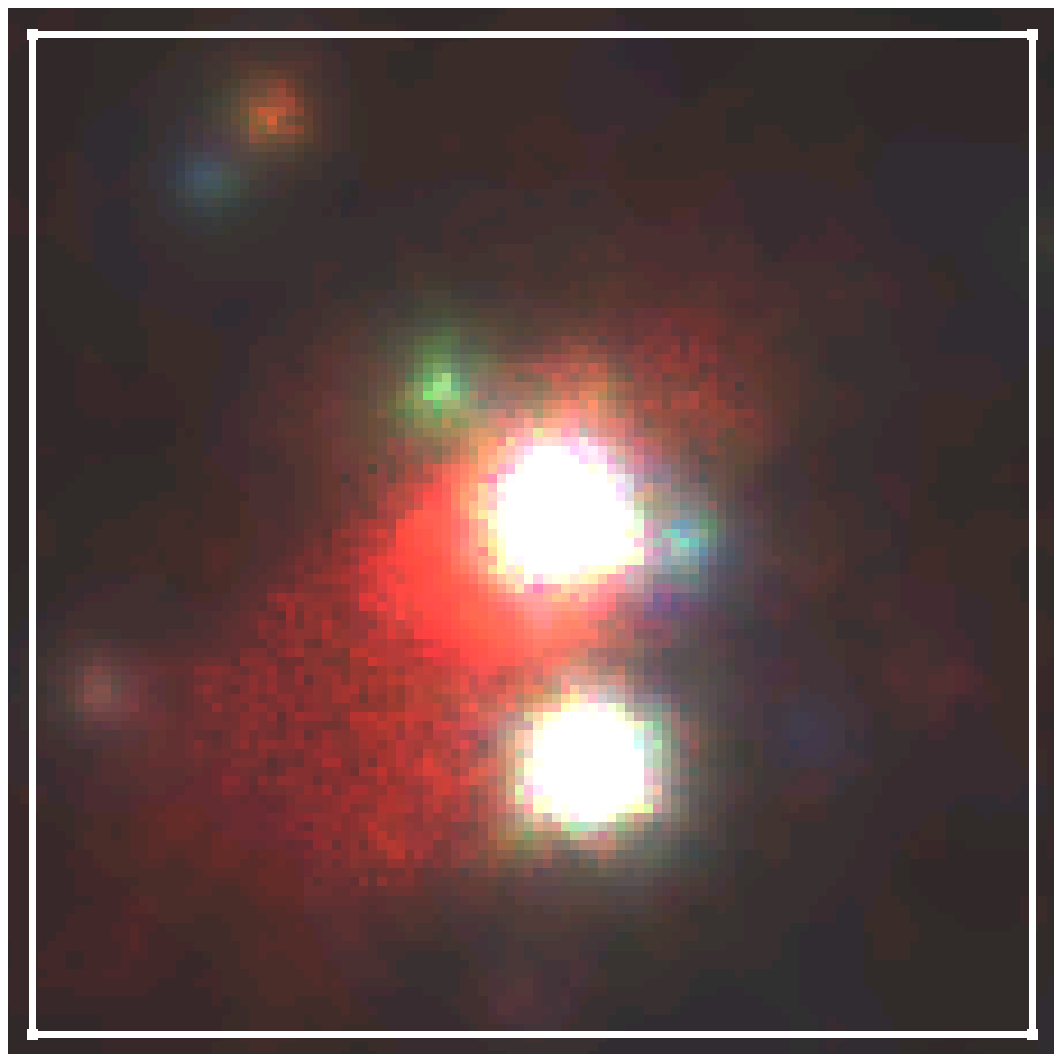}}
\end{centering}
\caption{ {\em Upper panel}: Three-colour, combined EPIC, $\sim$30$'$$\times$ 30$'$ image from the 2003 XMM-Newton observation of NGC 253; the intensity scale is histogram equalised. North is up, East is left. The energy bands used were 0.3--2.0 keV (red), 2.0--4.0 keV (green) and 4.0--10.0 keV (blue). The white ellipse represents the V band  D$_{25}$ contour of NGC 253. {\em Lower panel}: Linear-scaled image of the central 2$'\times 2'$ region; several  point sources are visible, while the red smudge is a superwind}\label{3im} 
\end{figure}

Removing intervals of high background  left good times of 45 ks in the pn and 69 ks in the MOS. We combined the cleaned MOS and pn images, and ran the  source detection algorithm provided with the XMM-Newton data analysis suite SAS version 7.0, producing a list of 185 sources that had no other sources within 10$''$.  
In Fig.~\ref{3im} we present a three-colour, combined EPIC image ($\sim 30'\times 30'$) of NGC253.  North is up, East is left. The white ellipse represents the V band D$_{25}$ isophot for NGC 253. We see that NGC 253 covers $\sim$20\% of the field; the vast majority of sources outside this region are likely to be background AGNs and foreground stars, but a few could be associated with globular clusters belonging to NGC 253. Both populations of X-ray sources, inside and outside the D$_{25}$ isophot, clearly display a wide range of colours. { We also show a linearly-scaled close-up of the central 2$'\times 2'$ region, showing several point sources in a region that looks like one unresolved source in the main image.}

\begin{table*}[!ht]
  \caption{Summary of emission models used when converting from 0.3--10 keV intensity to flux for an on-axis source with 15$''$ extraction radius. The model description, line-of-sight absorption, spectral index are given in columns 1--3. Column 4 shows the unabsorbed pn flux equivalent to 1 count s$^{-1}$ in the 0.3--10 keV band for an on-axis source with 15$''$ extraction radius}
\begin{tabular}{rccc}
  \hline
  \noalign{\smallskip}
Model & $n_{\rm H}$ / 10$^{22}$ atom cm$^{-2}$ & $\Gamma$ & $F_{0.3-10 keV}^{1 ct/s}$ / 10$^{-15}$ cm$^{-2}$ erg s$^{-1}$\\
\noalign{\smallskip}
\hline
\noalign{\smallskip}
 Model I: Standard model  & 0.013 & 1.7 & 4006\\
Model II: Best fit to all NGC253 sources (Total) & 0.34 & 2.49 & 8252 \\
Model III: Total$-$ULX  & 0.19 & 2.40 & 5365\\
Model III: ULX & 0.5 & 2.65 & 11693\\
 \noalign{\smallskip}
\hline
\noalign{\smallskip}
\end{tabular}
\end{table*}

For every source, we obtained an extraction region with radius 12--40$''$. In general, we used radius of 20$''$, { except for  sources with large PSFs due to high off-axis angle, where a 40$''$ radius was used}, or in very crowded regions, where a radius of 12--15$''$ was used.

We also created a corresponding background region for every source. We required that the background be on the same CCD as the source for all three EPIC cameras, that there be no point sources in the background, and that its intensity per unit area be smaller than for the source region. The resulting background regions had areas 1--35 times greater than their corresponding source regions; for 75\% of sources, the background : source area ratio $>$ 3.

We extracted pn and MOS  source and background spectra in the 0.3--10 keV range, combining the MOS1 and MOS2 spectra if the source was present in both cameras.
We obtained fits to 140 sources   with $>$50 source counts in  the pn { and/or} MOS spectra, using power law, blackbody and bremsstrahlung models; all models included line-of-sight absorption. We considered all fits with null hypothesis probability $\geq$0.05 as acceptable. If none of these spectral models provided a good fit, we considered a two-component model consisting of a power law and blackbody, as seen in Galactic X-ray binaries. We used the best fit model to obtain a 0.3-10 keV, unabsorbed flux for each source.

\section{Fluxes from different models}
For these 140 sources, we obtained fluxes from the source intensities using some of the methods employed when creating the XLFs used by G03. B07 describes this comparison in detail, but here we consider luminosities obtained using: (I) a standard emission model with Galactic line-of-sight absorption, (II) the best fit power law model to the summed spectrum of all sources within the D$_{25}$ of NGCc253 (III) best summed spectra of NGC253 point sources divided into ULXs and non-ULXs.

When converting from intensity to luminosity using particular emission models, we corrected the observed intensities for background, vignetting and encircled energy. Hence, we obtained for each source the intensity of an equivalent on-axis source with a 15$''$ extraction radius. We then used XSPEC to find the conversion  factor for each model, i.e. the unabsorbed flux required to give 1 count s$^{-1}$ in the 0.3--10 keV  band for an on-axis source with 15$''$  extraction radius. The parameters of these models are summarised in Table 1, along with their conversion factor. We immediately see that luminosities obtained from these different methods vary by a factor of 3 for the ULXs and a factor of 2.1 for the normal population. We also note that the major influence on the conversion factor appears to be the absorption, which is 20--50 times higher for the best fit models than for the standard model assuming Galactic absorption.

\begin{figure}
\resizebox{\hsize}{!}{\includegraphics[angle=270,scale=0.6]{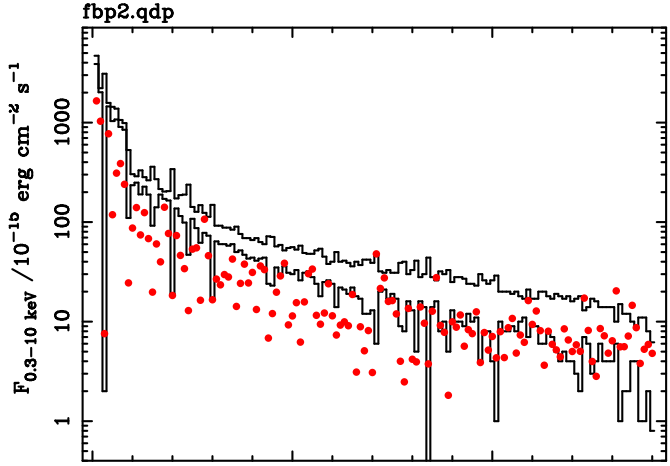}}
\caption{ Comparison of the standard model, i.e Model I, flux (filled circles) with the 90\% upper and lower flux limits from the best fit models (stepped lines). The y-axis is log-scaled, while the x-axis serves only to separate the sources. Models II and III differ from Model I only in scaling factor. }\label{sbc} 
\end{figure}

 In Fig.~\ref{sbc}, we compare the 90\% upper and lower flux limits from the best fit to each of the 140 bright sources with $>$50 counts in the pn and/or MOS spectra  with the flux obtained from assuming a standard model (Model I). The flux limits are represented by stepped lines, while the points represent PIMMS fluxes.   The x-axis is not numbered as it  serves only to separate the sources; the sources were ordered by decreasing best fit flux for clarity. The y-axis is log-scaled and shows the 0.3--10 keV flux. The standard model fluxes are consistent with the individually fitted fluxes for only $\sim$30\% of the sources; the fluxes agree more at lower fluxes, but the standard model systematically underestimates the individually fitted fluxes for the brighter sources. This is probably because brighter X-ray binaries are systematically softer, and hence the XMM intensity to flux correspondence is non-linear.

 The total flux from freely modelling spectra of all sources within the D$_{25}$ of NGC253 excluding the nuclear region is 2.8 times higher than the Model I flux for the same sources; the freely modelled total flux is 40\% higher than for Model II, and 10\% higher than for Model III. It is unsurprising that Model III is most successful at reproducing the freely fit luminosity, as it employs two conversion factors to NGC253 sources grouped by luminosity, rather than a single conversion factor; hence, Model III best samples the non-linear conversion between flux and intensity. Subdivision of the X-ray population into intensity bands, and obtaining corresponding conversion factors (like Roberts et al., 2002), is therefore likely to be the best approach to fitting low-photon data.

\section{Luminosity functions of the NGC 253 population and background}
\label{lf}

\begin{figure}
\resizebox{\hsize}{!}{\includegraphics[angle=270,scale=0.6]{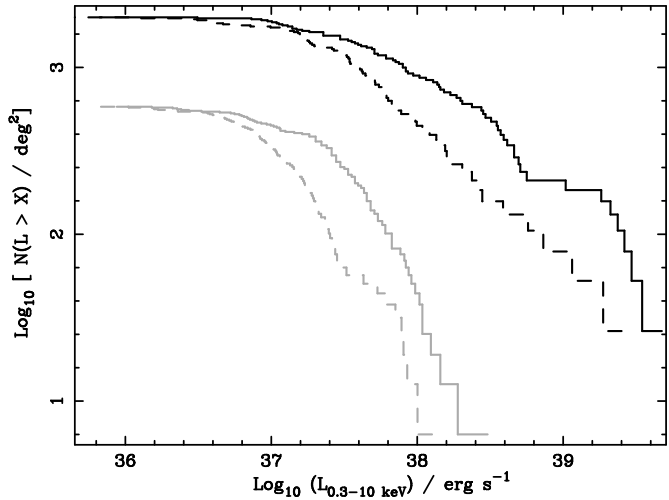}}
\caption{ X-ray luminosity functions (XLFs) for the NGC253 (black) and background (grey) populations, assuming a distance of 4 Mpc. Solid XLF are derived from best fit models, while dashed XLFs assume a standard model. The nuclear source  and QSOs  are removed from the NGC253 XLF, and a foreground star  is removed from the background XLF because its best fit luminosity has 100\% uncertainties. Kolmogorov-Smirnov (K-S) testing shows that the NGC253 and background populations have a probability of 1.8$\times$10$^{-7}$ for being drawn from the same population.}\label{lf1_2} 
\end{figure}

In Fig.~\ref{lf1_2} we show  XLFs inside and outside the D$_{25}$ region of NGC253,  for the 140 sources bright enough to fit, excluding the nuclear region and { two AGNs from the NGC253 population, and the foreground star from the background population}; the 0.3--10 keV  luminosity is plotted on the x-axis, and the number of sources per square degree  with higher fluxes given on the y-axis. The D$_{\rm 25}$ region and background region cover 137 and 570 square arcminutes respectively, assuming a circular field of view with 15$'$ radius. The black lines represent the NGC253 XLFs, while the grey lines represent the background XLFs; solid lines are XLFs derived from individually fit   models, and dashed lines are XLFs derived from the standard model . The luminosities given assume a distance of 4 Mpc, although this is unlikely to apply for many of the background sources. A Kolmogorov-Smirnoff test shows that the NGC253 and background populations have a probability of 1.8$\times$10$^{-7}$ for being drawn from the same population.

We first note that the spatial density of X-ray sources for the NGC253 population
 is a factor of $\sim$4 times higher than that of the background population; this is in keeping with the idea that the NGC253 population mainly belongs to NGC 253, while the background population consists mainly of background or foreground objects. 

{
We next note that for NGC253, the   XLF derived from individual spectral fits is considerably flatter than the XLF obtained using the standard model (Model I). We used the SHERPA fitting software to model these XLFs with broken power law models, using Cash statistics to obtain uncertainties. The goodness of fit of the best fit broken power law models were then tested using a method adopted from that of Crawford, Jauncey \& Murdoch (1970), who developed a  goodness of fit estimation for modelling unbinned  data such as lumiosity functions. The best  IF NGC253 model consists of a broken power law that changes spectral index from 0.138$\pm$0.016 to 0.62$\pm$0.06 at a 0.3--10 keV luminosity of 4.6$\pm$0.9$\times$10$^{37}$ erg s$^{-2}$; the good fit probability was 0.14. The best  SM NGC253 model breaks from 0.11$\pm$0.03 to 0.81$\pm$0.09 at 2.3$\pm$0.3$\times$10$^{37}$ erg s$^{-1}$, with a good fit probability of 0.19. We note that luminosities lower than the break are likely artefacts of incompleteness, and are not particularly meaningful. We applied the best fit SM model to the IF XLF, and found that the good fit probability was only 0.003; hence the differences between the XLFs are statistically significant. The higher conversion factors of Models II and III with respect to Model I mean that their XLFs are correspondingly flatter.

}

\section{Conclusions}

Grimm et al. (2003) report a Universal HMXB XLF derived from published XLFs of several nearby galaxies. They also derive relations between the star formation rate and (i) the total luminosity of the point X-ray sources in the galaxies and (ii) the number of X-ray sources in a galaxy with 2--10 keV luminosity $>$2$\times$10$^{38}$ erg s$^{-1}$.  However, the published XLFs were produced using a number of methods, in most cases assuming Galactic line-of-sight absorption. 

We have tested several of these models using a deep XMM-Newton observation of the nearby galaxy NGC253, included in a secondary sample of Grimm et al. (2003). We obtained freely modelled luminosities for the 140 brightest sources in the field and also obtained the conversion factors from intensity to flux for some of these different models. We  found them to vary by a factor of $\sim$3. We found the biggest influence on the conversion factor to be the absorption; Model I assumed Galactic line-of-sight absorption, while absorptions 20--50 times higher than this were obtained for the other methods. Since the universal XLF and relations between SFR and X-ray properties were obtained using a mixture of methods, we find them to be inconclusive. 

We have also found from freely fitting the spectra of 140 bright sources that the corresponence between count-rate and flux is non-linear; this is largely due to the systematic softening of the spectra of more luminous sources. Hence it is unwise to employ a single emission model when describing the X-ray populations of nearby galaxies. Ideally, one would construct XLFs only from luminosities derived from free spectral modelling. To get the most out of the low-photon data, we recommend the stacking method of e.g. Roberts et al. (2002).

\begin{acknowledgements}
Astronomy at the Open University is funded by a STFC (formerly PPARC) Rolling Grant.
\end{acknowledgements}

\end{document}